\documentclass[conference]{IEEEtran}
\IEEEoverridecommandlockouts

\usepackage{cite}
\usepackage{amsmath,amssymb,amsfonts}
\usepackage{algorithm}
\usepackage{algpseudocode}

\usepackage{graphicx}
\usepackage{textcomp}
\usepackage{xcolor}
\usepackage{multirow}
\usepackage{makecell}
\usepackage{float}
\usepackage{lipsum}

\begin{document}

\title{
Identification of Sub/Super-Synchronous Control Interaction Paths Using Dissipative Energy Flow
\\
\thanks{This work is supported by the Energy System Co-Design with Multiple Objectives and Power Electronics
    (E‑COMP) Initiative at PNNL (Contract DE‑AC05‑76RL01830).}%
}
\author{\IEEEauthorblockN{Kaustav Chatterjee, Sameer Nekkalapu, Sayak Mukherjee, Ramij R. Hossain, and Marcelo Elizondo\\
\emph{Pacific Northwest National Laboratory, Richland, WA, USA} 
}}
\maketitle  

\begin{abstract}
Sub- and super-synchronous control interactions (SSCIs) are oscillations arising from adverse interactions between inverter-based resource (IBR) controls and the power network. SSCIs often involve multiple frequencies and propagate through complex, interconnected paths, making it difficult for model-based approaches to identify both the sources and the paths of oscillatory energy flow. This paper extends the Dissipative Energy Flow (DEF) method, originally developed for low-frequency electromechanical oscillations, to identify SSCI sources and dynamic interaction paths across multiple frequencies using three-phase voltage and current measurements. The approach operates in the $dq$ frame using dynamic phasors, enabling mode-specific DEF computation from bandpass-filtered signals. An electromagnetic transient (EMT) case study on a meshed network with synchronous generator and type-3 wind farm resources under series-compensated conditions demonstrates the method’s capability to distinguish frequency-dependent source and sink roles, including cases where the same resource acts as a source at one frequency and a sink at another. The results show DEF can provide a physics-based and automation-friendly tool for SSCI diagnosis in IBR-rich grids.
\end{abstract}


\begin{section} {Introduction}
Sub- and super-synchronous control interactions (SSCIs) are sustained oscillations caused by adverse interactions between the fast control loops of inverter-based resources (IBRs) and the network at frequencies below or above the nominal synchronous frequency \cite{stability_def, cheng2022real}. They form a subset of sub- and super-synchronous oscillations (SSOs)  that are purely control-driven and do not involve mechanical resonance \cite{stability_def}. While historically associated with wind farms connected through series-compensated transmission lines, SSCIs have also been observed in uncompensated and weak grids \cite{cheng2022real}, such as 4 Hz voltage-control oscillations in ERCOT’s Texas grid \cite{Texassystem_Mainpaper} and 3.5 Hz real- and reactive-power oscillations in Hydro One \cite{li2019asset}. These interactions can accelerate equipment degradation, trigger protection misoperations, and compromise stability, motivating mitigation through converter retuning, supplemental damping controls, and coordinated system design.

Tracing the paths of oscillatory energy exchange between components can provide valuable insight into how targeted control adjustments may mitigate SSCIs. However, identifying the source and propagation routes of SSCIs is challenging due to the complex nature of the interactions. Unlike forced oscillations (FOs), where a single component can often be uniquely identified as the dominant contributor, SSCIs may involve multiple components—sometimes at more than one frequency—exchanging energy through several interconnected pathways. To address the source localization problem, several analytical methods have been proposed, among which the Dissipative Energy Flow (DEF) method \cite{chen_def1, MASLENNIKOV201755} stands out for its intuitive, physics-based foundation and straightforward implementation. Originally developed for identifying the sources of low-frequency electromechanical oscillations in synchronous generator–dominated grids, the DEF framework can be extended to higher-frequency phenomena and adapted for networks with high IBR penetration \cite{Fan2023OscillationSourceDetection}. Prior studies have examined the theoretical extension of DEF to inverter-based generation resources, as well as inverter-based transmission assets such as STATCOMs and VSC-HVDC systems \cite{chatterjee_statcom, chatterjee_hvdc}. However, these works have primarily focused on low-frequency oscillations and have not addressed the unique challenges associated with SSCIs. 

A recent study \cite{Estevez2025CDEFExtension} applied an enhanced formulation of DEF, referred to as complex-DEF (CDEF) \cite{cdef}, to address the SSO source localization problem in an IEEE test system. While CDEF demonstrates improved localization accuracy, it requires careful tuning of a hyper-parameter $K$ that can vary with network conditions and oscillation frequency, making it less practical for real-time or wide-scale deployment. 

In this paper, we propose an alternate approach that builds on the original DEF formulation, computing the energy functions for the SSO components from dynamic phasors in the $dq$ frame, derived from three-phase voltage and current measurements at the point of interconnection of the resource. The proposed method preserves the simplicity and parameter-free nature of the original DEF, improving robustness and suitability for online implementation. Furthermore, this work extends the application of DEF to multi-frequency SSOs, demonstrating cases where the same resource acts as a source at one frequency and a sink at another—a scenario not studied in the CDEF-SSO paper \cite{Estevez2025CDEFExtension}. The proposed method is demonstrated on an electromagnetic transients model of a meshed test network with synchronous and DFIG-based wind generation under varying series-compensation levels, illustrating accurate source-path localization without case-specific calibration.

The paper is organized as follows. Section II reviews the background on SSCIs, outlining the underlying mechanisms and typical interaction paths, and presents a simulation study of a real-world event to motivate the problem. Section III introduces the DEF framework and details the proposed method for mapping SSCI interaction paths. Section IV applies the method to a synthetic test system to assess its effectiveness, and Section V summarizes the key findings and conclusions.
\end{section}

\begin{section}{Sub/Super-Synchronous Control Interactions}
The IEEE PES IBR SSO Task Force categorizes SSCIs into two types \cite{cheng2022real}: (1) {series-compensator} SSCI, caused by interactions between IBR control dynamics and the resonant modes of series-compensated lines, and (2) {weak-grid} SSCI, driven by control–grid impedance interactions under low short-circuit strength. The former is resonance-driven, typically when the series-capacitor resonant frequency lies within the IBR control bandwidth; the latter is control-instability-driven and can occur without series compensation. Both have been observed in practice \cite{fan2022real}, producing sustained or growing oscillations that propagate across network paths. 


In series-compensator SSCI, instability occurs when the resonant frequency of the compensated network aligns with the bandwidth of the inverter’s outer loops (active/reactive power or voltage regulators). High PI gains and PLL interactions can introduce negative damping, allowing transient energy to circulate between the converter controls, inner current loop, PLL, and the series-compensated network \cite{cheng2022real}.
This adverse coupling between control loops (outer, inner, and PLL) and network impedance enables oscillatory energy exchange across sub- or super-synchronous frequencies. A simulation study of a real-world series compensator SSCI event is presented next to illustrate the mechanisms and interaction paths.

\subsection*{Simulation Study of a Real-World Event}
In 2009, the southern Texas grid experienced a sub-synchronous control interaction event involving a large wind farm equipped with doubly-fed induction generators (DFIGs) \cite{Texassystem_Mainpaper}, \cite{Texassystem_Suggestedpaper}. Following the tripping of a 345 kV line to clear a fault, the plant became radially connected through a 50\% series-compensated line. The interaction between the DFIG control loops and the series-capacitor resonance excited sub-synchronous oscillations, causing significant concerns in the system.

\begin{figure}[h]
    \centering \vspace{-0.2cm}
    \includegraphics[width=1.01\linewidth]{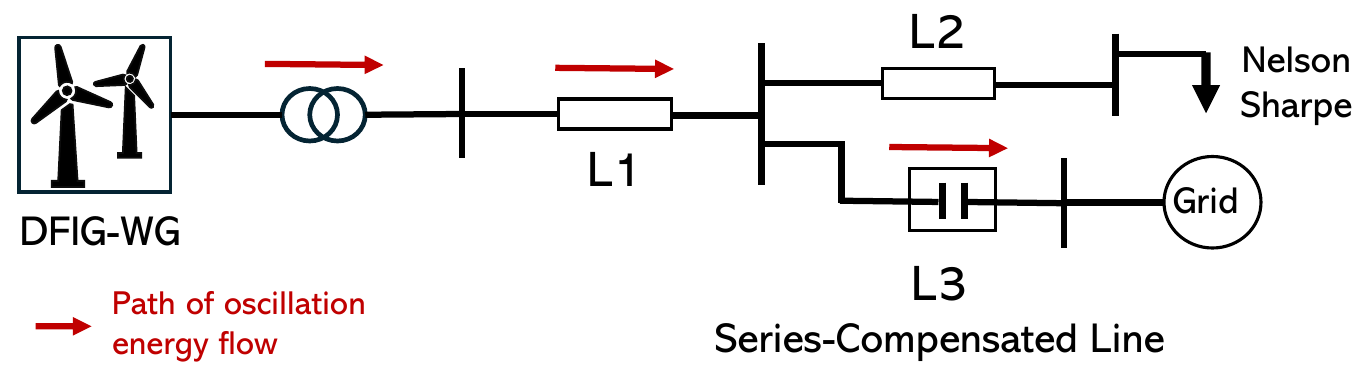}
    \caption{Single-line diagram of the test system, used to replicate the SSCI event from Texas, showing the path of oscillation energy flow or control interaction.} \vspace{-0.1cm}
    \label{fig:Circuit_Texas}
\end{figure}

This work replicates the operating condition, in the electromagnetic transients simulation platform PSCAD, using the network shown in Fig. \ref{fig:Circuit_Texas}. In this, around 200 MW of generation
is being produced by the DFIG wind source. The load at the Nelson Sharpe bus has been considered to be negligible to mimic the tripping of the 345 
kV line. Series compensation is set to 400 µF, selected from sensitivity studies as the minimum value that remains stable under nominal conditions but becomes unstable with small changes to control gains in the DFIG. A parallel breaker provides soft capacitor switching to mitigate overvoltage. Transmission lines L1, L2, and L3, between Zorillo-Ajio, Ajio-Nelson Sharpe, and Ajio-Rio Hondo locations, are modeled using lumped 60 Hz PI-coupled parameters. In the simulated event (Figs. \ref{fig:Currents_WG_Texas}), the proportional gain $K_p$ of the two inner loop and the two outer-loop rotor side controls of the DFIG is increased from 2 to 4 at $t = 15 $s. The gain adjustment excites two sideband oscillatory components at 12 Hz and 108 Hz in the instantaneous phase currents, along with their harmonics, arising from the DFIG–series capacitor interaction. These components also appear in the $dq$-frame as a shifted frequency of $60-12 = 48$ Hz (equivalently $108-60 = 48$ Hz), as shown in Fig.~\ref{fig:psd_WG_Texas}. In this scenario, the source of instability is the DFIG’s outer-loop control interacting with the network resonance path formed by the series-compensated line, with oscillatory energy circulating between the converter controls, the inner current loop, and the resonant network branch via lines L1 and L3 (Fig.~\ref{fig:Circuit_Texas}). The corresponding figures illustrate how this energy propagates along the interaction path involving the DFIG and the compensated line, sustaining the observed oscillations.

\begin{figure}[h]
    \centering \vspace{-0.2cm}
    \includegraphics[width=1\linewidth]{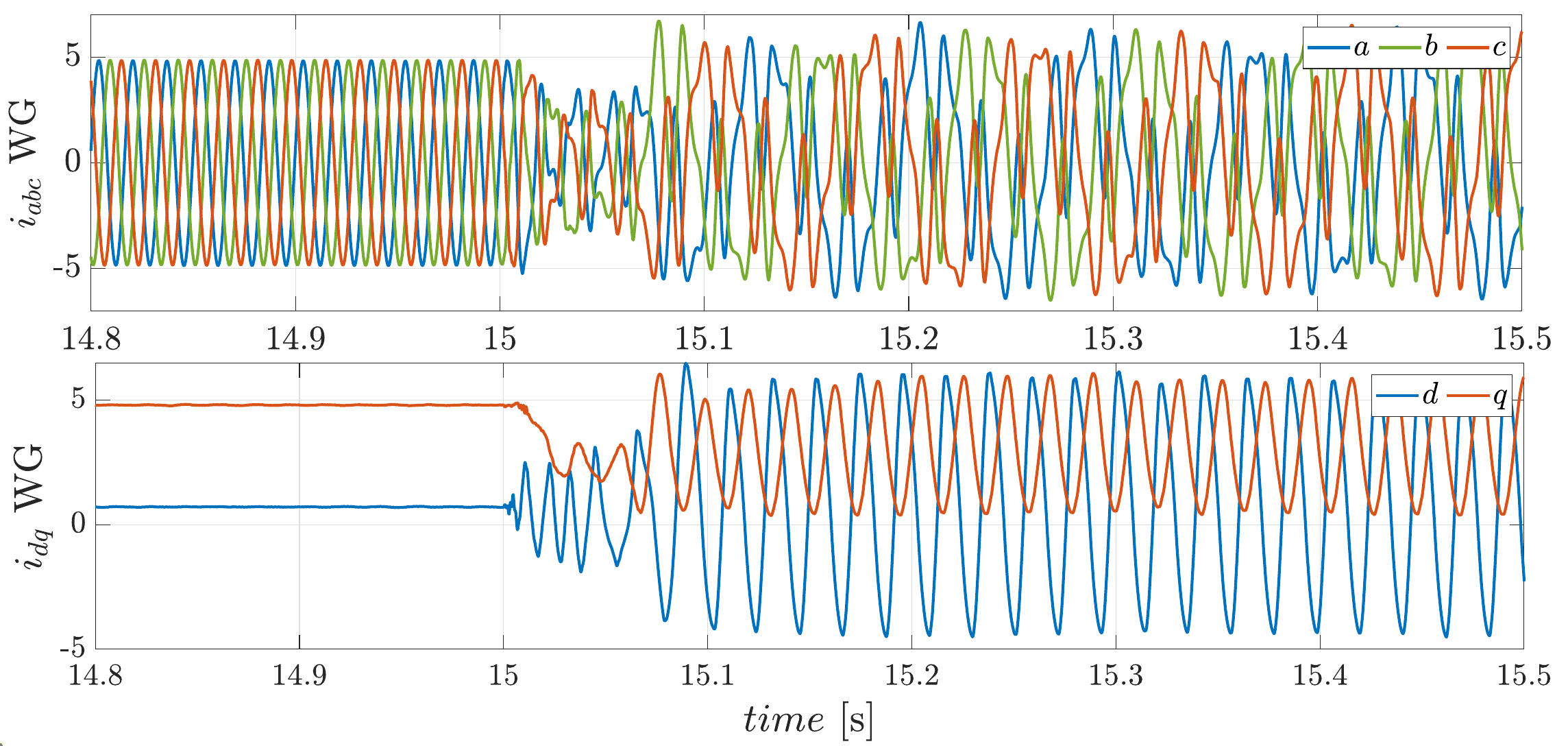} \vspace{-0.5cm}
    \caption{$abc$ phase waveforms and $dq$ components of the current injected by the WG in the Texas system.}
    \vspace{-0.4cm}
    \label{fig:Currents_WG_Texas}
\end{figure}
\begin{figure}[h]
    \centering\vspace{-0.1cm}
    \includegraphics[width=1\linewidth]{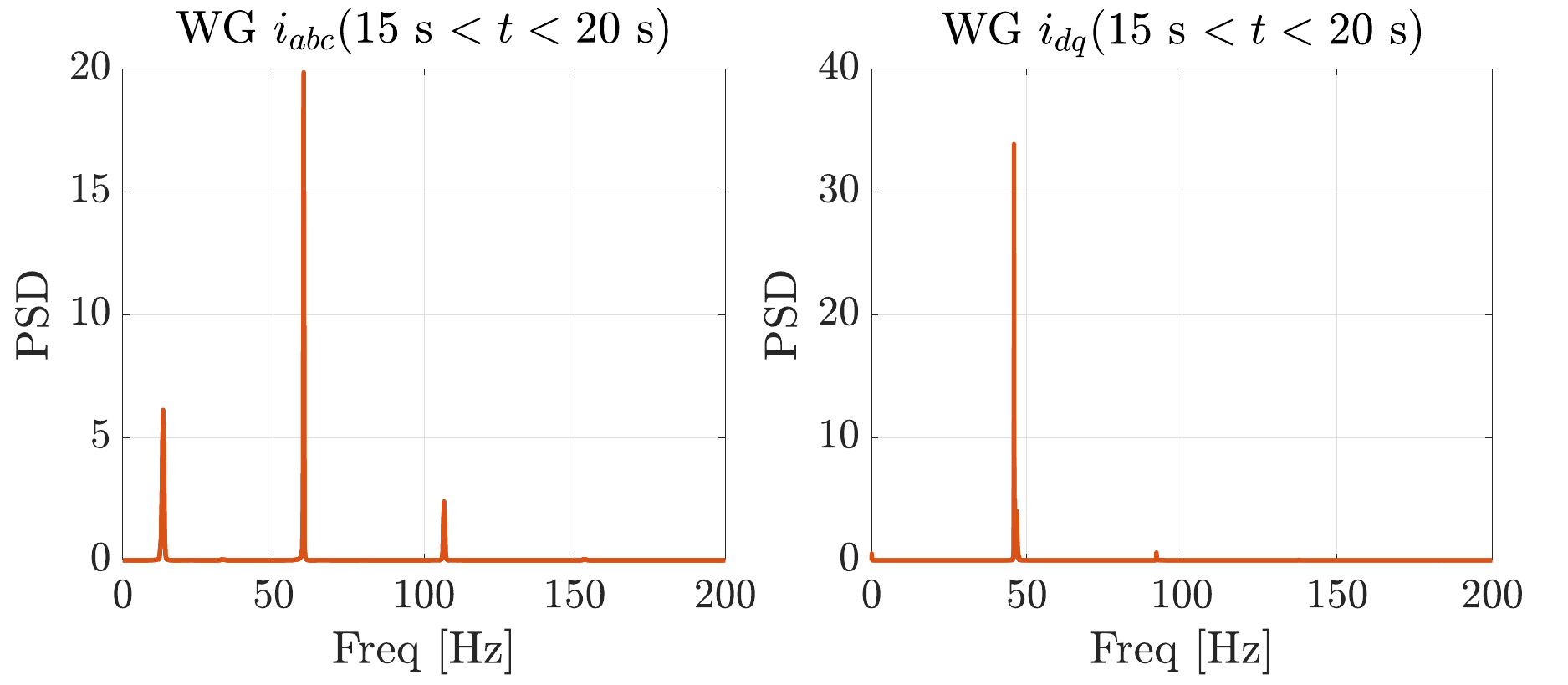}
    \caption{Frequency spectrum of the currents injected by the WG computed for (a) instantaneous three-phase components and (b) $dq$ components. }
    \vspace{-0.2cm}
    \label{fig:psd_WG_Texas}
\end{figure}

This example demonstrates how SSCI can be traced to a specific source—the DFIG outer-loop controls—and a well-defined interaction path through the network impedance. Such events underscore the need for systematic methods to localize instability sources and map energy propagation routes. The following section introduces the DEF framework, a physics-based, automation-friendly approach for identifying SSCI sources and interaction paths from measurement data.

\end{section}

\begin{section}{Determination of Interaction Path Using Dissipative Energy Flow}

Dissipative Energy Flow (DEF) formulations proposed in the literature for source localization of low-frequency oscillations have predominantly utilized positive-sequence phasor representations, as outlined below. 
\begin{equation} \label{DEF} \small
    W \, = \, \int \Im \{ \vec{I}^*\,d\vec{V}\} \, = \, \int P d\theta + \frac{Q}{V} dV
\end{equation}
In (\ref{DEF}), the quantity $W$ denotes the transient oscillation energy injected into the power network from a node with voltage $\vec{V}$ and current injection $\vec{I}$. The terms $P$ and $Q$ are, respectively, the real and reactive powers injected from that node. The time-derivative of $W$ averaged over a cycle $\tau$ -- referred to as the slope of DEF and denoted by $\dot{W}$, determines the rate of dissipation of transient energy emanating from the node. If the slope of the DEF from a node $\dot{W} > 0$, the shunt element connected to that node is characterized as an \emph{oscillation source}. Alternatively,  if $\dot{W} < 0$, the shunt element is characterized as an \emph{oscillation sink}.

\subsection{DEF Formulation in the $dq$ Frame}
While fundamental-frequency phasor representations are adequate for analyzing low-frequency electromechanical oscillations, they are insufficient for studying higher-frequency transients such as those associated with sub-synchronous oscillations (SSOs). To overcome this limitation, we adopt a dynamic phasor framework for DEF, employing $dq$-sequence components derived from $abc$ quantities via the Park transformation, as follows:
\begin{equation}
    \label{DEF_dq}
    \begin{aligned}
        W \, = \, \int \Im \{ \vec{I}^*\,d\vec{V}\} \, &= \, \int \Im \big\{ (i_d - j\,i_q) \, (d v_d + j \,d v_q)\big\} \\
        &= \, \int i_d \,dv_q - i_q \,dv_d
    \end{aligned}
\end{equation}
In this approach, three-phase ($abc$) voltage and current data, obtained either from field-deployed point-on-wave (POW) recorders or from electromagnetic transient (EMT) simulations, are transformed into their corresponding $d$ and $q$ components. In the original three-phase waveform data, transient oscillations appear as sidebands around the fundamental frequency component. When transformed into the 
$dq$ frame, these sideband oscillations manifest as oscillations at the difference frequency — the frequency difference between the fundamental and the sidebands. For example, sideband components at 
$f_0 \pm \Delta f$ in the $abc$ frame translate into oscillations at a frequency of $\Delta f$ in the $dq$ components. 

\begin{algorithm}[t]
\caption{DEF-Based SSCI Path Determination}
\begin{algorithmic}[1]
\Require $v_{abc}(t), i_{abc}(t)$ at all nodes and incident lines; sampling $f_s$; slope thresholds $\epsilon$ and $\epsilon_n$
\State $(v_d,v_q,i_d,i_q) \gets \text{ParkTransform}(v_{abc}, i_{abc})$
\State {Calculate dominant SSO mode frequencies $\{ f_m\}$: $\{ f_m\} \gets \text{WindowedSpectralID}(v_d,v_q,i_d,i_q)$}
\ForAll{$\Delta f_m$}
    \State {Mode-specific bandpass filtering centered at $\Delta f_m$} $~~~~(\Delta v_d,\Delta v_q,\Delta i_d,\Delta i_q) \gets \text{Bandpass}_{f_m}(v_d,v_q,i_d,i_q)$
    
    \State Initialize directed graph $\mathcal{G}_m = (\mathcal{N}, \emptyset)$
    \ForAll{edges $e \equiv \text{line}(i \leftrightarrow j) \in \mathcal{E}$}
        \State $W_e[0] \gets 0$
        \For{$k=0$ to $K-1$}
            \State $W_e[k{+}1] \gets W_e[k] + \Delta i_{d,e}^{(i)}[k]\big(\Delta v_q^{(i)}[k{+}1]- \Delta v_q^{(i)}[k]\big)$
            $- \;\Delta i_{q,e}^{(i)}[k]\big(\Delta v_d^{(i)}[k{+}1]-\Delta v_d^{(i)}[k]\big)$  
        \EndFor
        \State $\dot W_e \gets \text{Slope}(W_e, T_\mathrm{win})$ 
        \Comment{$T_\mathrm{win}$: window length  }
        \If{$\dot W_e > \varepsilon$}
            \State Add directed edge $(i \rightarrow j)$ to $\mathcal{G}_m$
        \ElsIf{$\dot W_e < -\varepsilon$}
            \State Add directed edge $(j \rightarrow i)$ to $\mathcal{G}_m$
        \Else
            \State No edge added (below threshold)
        \EndIf
    \EndFor
    \State Characterization of nodes: for each $n \in \mathcal{N}$, set
    \[
    \mathcal{C}_{m,n} \gets 
    \begin{cases}
        \text{Source}, & \sum_{e \in \mathcal{L}_n} \dot W_e - \sum_{e \in \text{in}(n)} \dot W_e > \varepsilon_n \\
        \text{Sink}, & \sum_{e \in \mathcal{L}_n} \dot W_e - \sum_{e \in \text{out}(n)} \dot W_e < - \varepsilon_n \\
        \text{Neutral}, & \text{otherwise}
    \end{cases}
    \]
\EndFor
\State \Return $\{\mathcal{G}_m\}, \mathcal{C}_{m,n}$ \Comment{$\{\mathcal{G}_m\}$: mode-$m$ directed graph for SSCI and  $\mathcal{C}_{m,n}$: mode-$m$ source/sink characterization of node $n$}. 
\end{algorithmic} 
\end{algorithm} 

In the first step of SSO characterization and analysis, a windowed Fourier-based spectral identification algorithm is applied to the $dq$ components to identify the dominant frequencies present in the signal. Subsequently, bandpass filtering is performed on the $dq$ voltage and current components to isolate the oscillations of interest. The resulting filtered signals are then used in the DEF computation defined in (\ref{DEF_dq_discrete}). 
\begin{equation} 
\label{DEF_dq_discrete} 
\begin{aligned}
      W [k+1] \, = \, W[k] \, &+ \, \Delta i_d [k] \, (\Delta v_q [k+1] - \Delta v_q [k] ) \, \\&- \,  \Delta i_q [k] \, (\Delta v_d [k+1] - \Delta v_d [k] )
\end{aligned}
\end{equation}

The equation in (\ref{DEF_dq_discrete}) is a discretized representation of (\ref{DEF_dq}), where $\Delta i_d, \Delta i_q, \Delta v_d$ and $\Delta v_q$ are the detrended bandpass-filtered $dq$ components of the injected current and node voltage, and $k$ indexes the time instant. 

\vspace{-0.1 cm}
\subsection{Determination of Interaction Path}
To determine the direction of oscillation energy flow, the dissipative energy term $W$ is computed at each node in the system for each connected line, specifically in the direction of outward current flow. For example, at a node $i$ connected to a set of lines $\mathcal{L}_i = \{ \ell_j \, |\,  j \in 1, 2, \dots N_i\}$, the DEF $W_{i}^{j}$ is calculated for each line $\ell_j$ using the current $I_{j}$ flowing out of the node $i$ and the corresponding node voltage $V_i$. If the slope of $W_{i}^{j}$ is positive ($\dot{W}_{i}^{j} > 0$), the oscillation energy is flowing outward from the node $i$ along line $j$. Conversely, a negative slope ($\dot{W}_{i}^{j} < 0$), indicates energy flow toward the node $i$ from that line. An algorithmic summary of the DEF-based procedure for determining SSO interaction paths is provided in Algorithm 1. Next, we demonstrate the effectiveness of the proposed approach using a test system adapted from \cite{PSCAD_SSCI_Example}.

\end{section}

\section{Illustrative Case Study}

\begin{figure}[h!]
\centering
\includegraphics[width=0.65\linewidth]{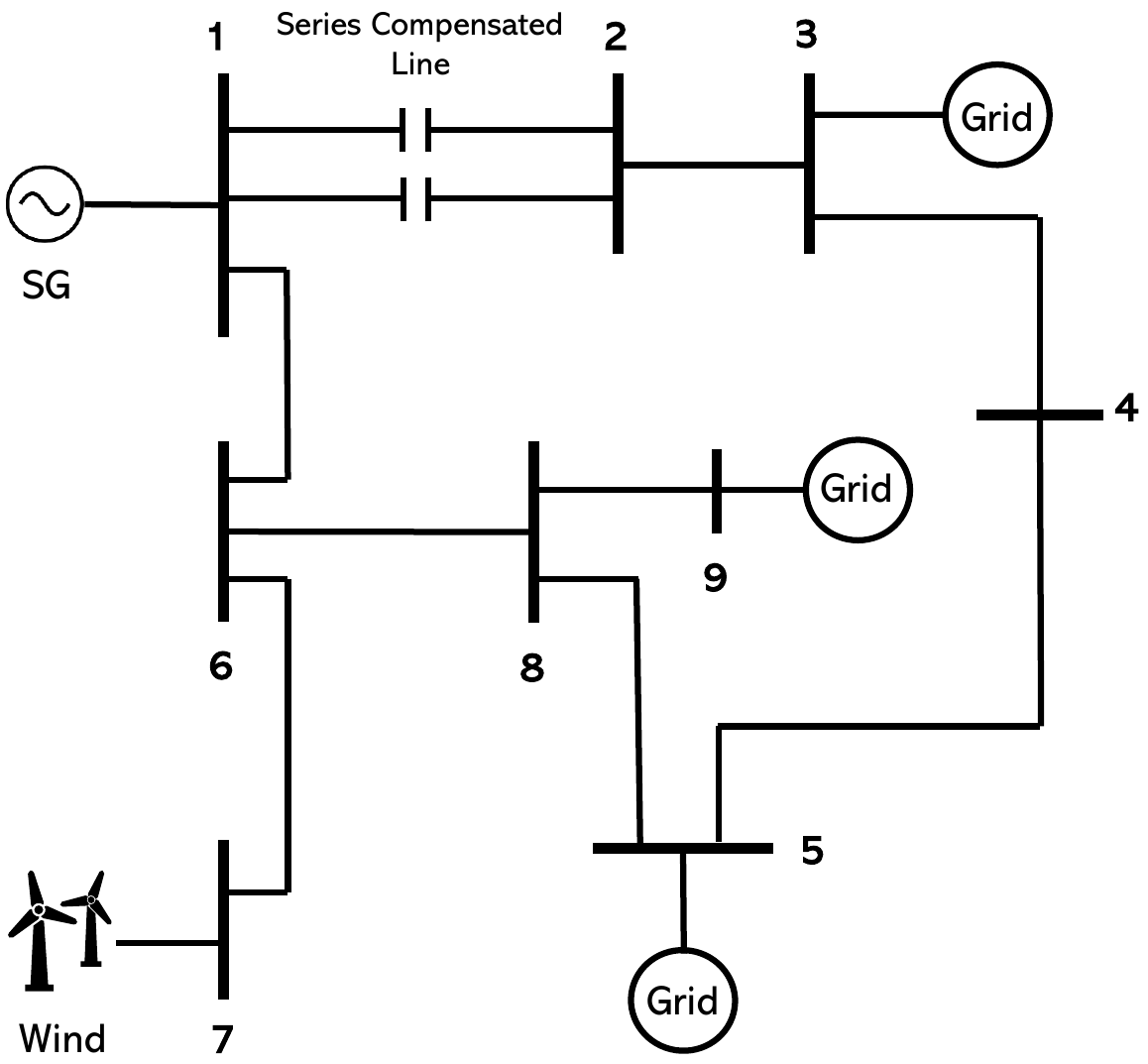}
\vspace{-0.1cm}
\caption{Test system with DFIG WG and SG interconnected to the grid.}
\label{fig:test_system}
\end{figure}

The test system from \cite{PSCAD_SSCI_Example} consists of a meshed network with a synchronous generator (SG) and a type-3 DFIG-based wind generator (WG), interconnected to a larger grid via both series-compensated and uncompensated long transmission lines, as shown in Fig.\ref{fig:test_system}. Under nominal operating conditions, the SG and WG supply 600MW and 200MW, respectively. The system’s dynamic behavior is studied, in PSCAD, for varying series-compensation levels, and it is observed that a capacitance of $25\mu\text{F}$ produces sustained oscillations.

The system is analyzed in two operating regimes: 
(1) \textit{Window-A} ($t=0$--$15$~s), where it operates with the selected compensation level and nominal SG and WG control parameters; and 
(2) \textit{Window-B} ($t=15$--$30$~s), where the proportional--integral (PI) gain in the DFIG-WG's outer real power control loop is detuned. 
The objective is to examine the control oscillations in each window and determine their corresponding interaction paths.

\begin{figure}[t!]
\centering \vspace{-0.2cm}
\includegraphics[width=\linewidth]{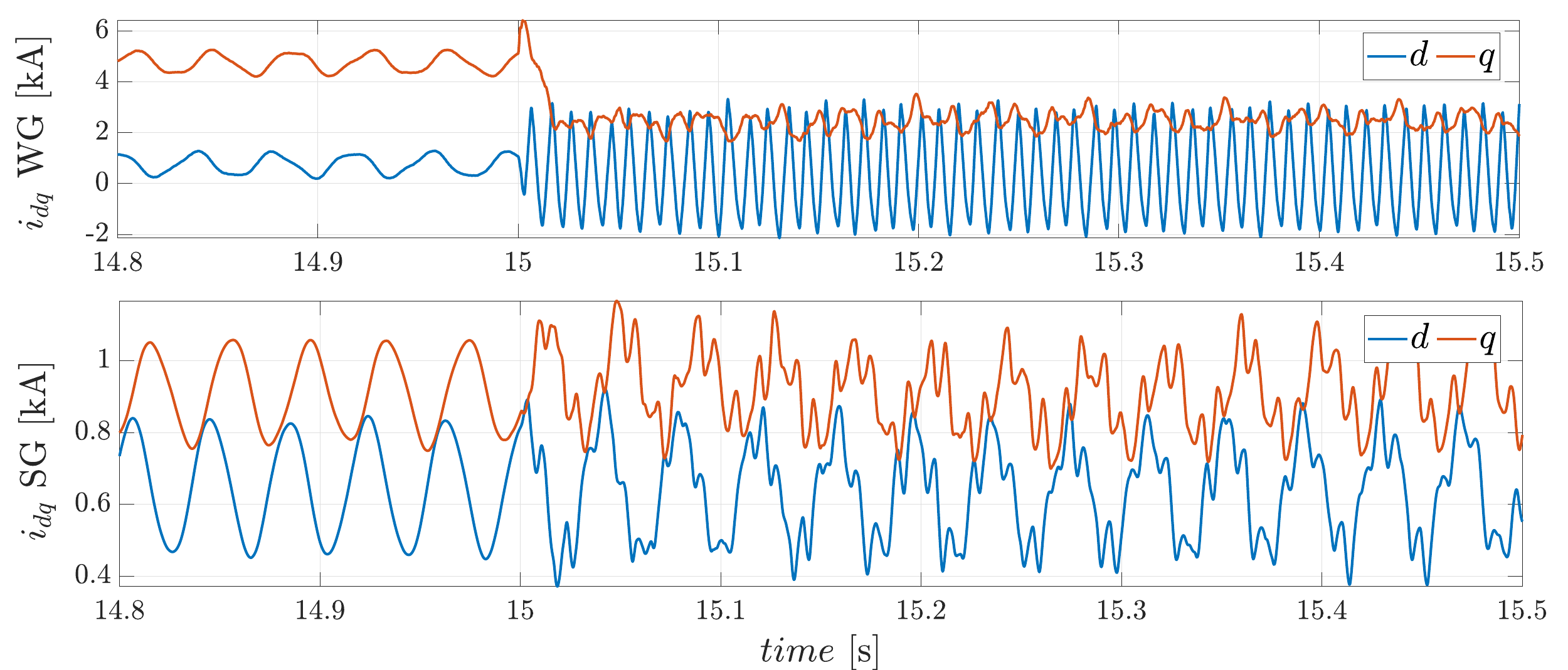}
\vspace{-0.7cm}
\caption{$d$- and $q$-axis components of the currents injected by the WG and SG derived from the three-phase quantities.}
\label{fig:idq_sg_wg} \vspace{-0.4cm}
\end{figure}

\begin{figure}[t!]
\centering
\includegraphics[width=\linewidth]{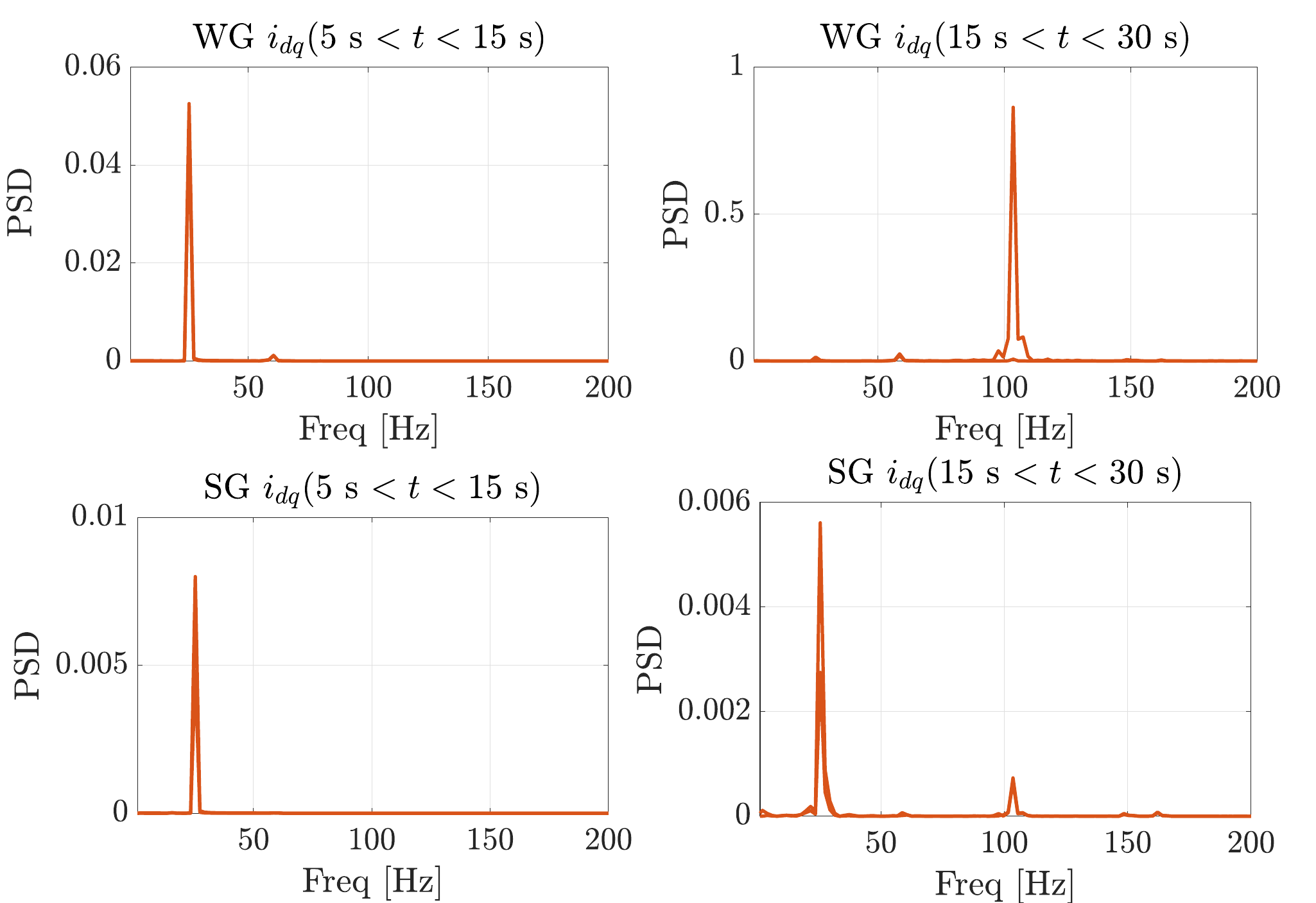}
\vspace{-0.7cm}
\caption{Frequency spectra (as PSD) of $dq$-axis currents injected by WG and SG in Windows A and B.} \vspace{-0.5cm}
\label{fig:psd_dq}
\end{figure}

The $d$- and $q$-axis current components injected by the WG and SG are shown in Fig. \ref{fig:idq_sg_wg}. The corresponding frequency spectra, computed using the power spectral density (PSD), are shown in Fig. \ref{fig:psd_dq}. In Window-A, a single 25Hz sub-synchronous component is present with contributions from both SG and WG. Following the PI-gain detuning, in Window-B, two oscillatory components: a sub-synchronous at 25Hz and another super-synchronous at 105Hz are observed, superimposed in the $dq$-domain currents (Fig. \ref{fig:idq_sg_wg}). The Dissipative Energy Flow (DEF) method is applied to identify oscillation energy sources, sinks, and interaction paths for each frequency component in its respective time window. Bandpass filtering is first applied to isolate the $dq$-axis currents at the target frequencies, after which DEF is computed as described in Algorithm~1.

\begin{figure}[b!]
\centering \vspace{-0.2cm}
\includegraphics[width=\linewidth]{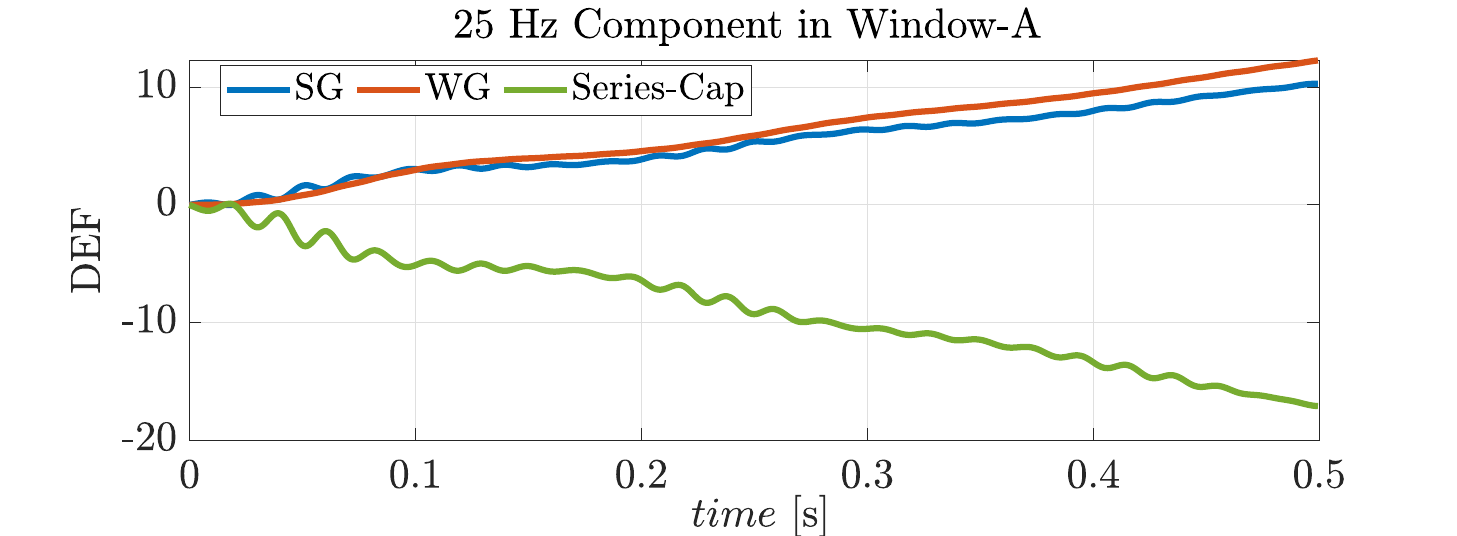}
\vspace{-0.7cm}
\caption{DEF from the grid elements into the grid computed from the data in Window-A between $14.5–15$ s for the 25 Hz component.}
\label{fig:def_25_before}
\end{figure}

\begin{figure}[b!]
\centering\vspace{-0.3cm}
\includegraphics[width=\linewidth]{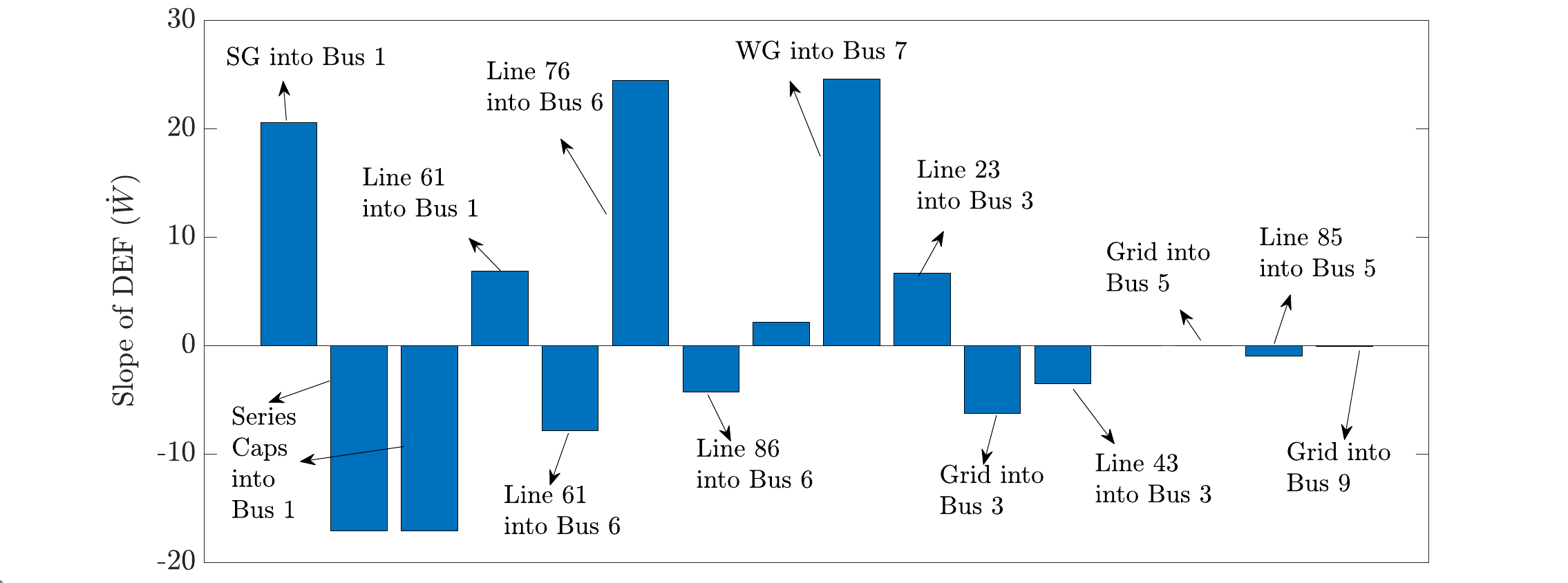}
\vspace{-0.3cm}
\caption{Average slope of DEF from different grid elements calculated for the 25 Hz in Window-A.}
\label{fig:slope_def_25_before}
\end{figure}

For Window-A, Fig. \ref{fig:def_25_before} shows the DEF at 25Hz injected by the SG, WG, and the series-compensated line into buses 1, 2, and 1, respectively. The average DEF slope $\dot{W}$ for each component, shown in Fig.~\ref{fig:slope_def_25_before}, indicates that both the SG and WG act as energy sources for this SSCI, with the series-compensated line acting as the primary sink.

\begin{figure}
\centering
\includegraphics[width=\linewidth]{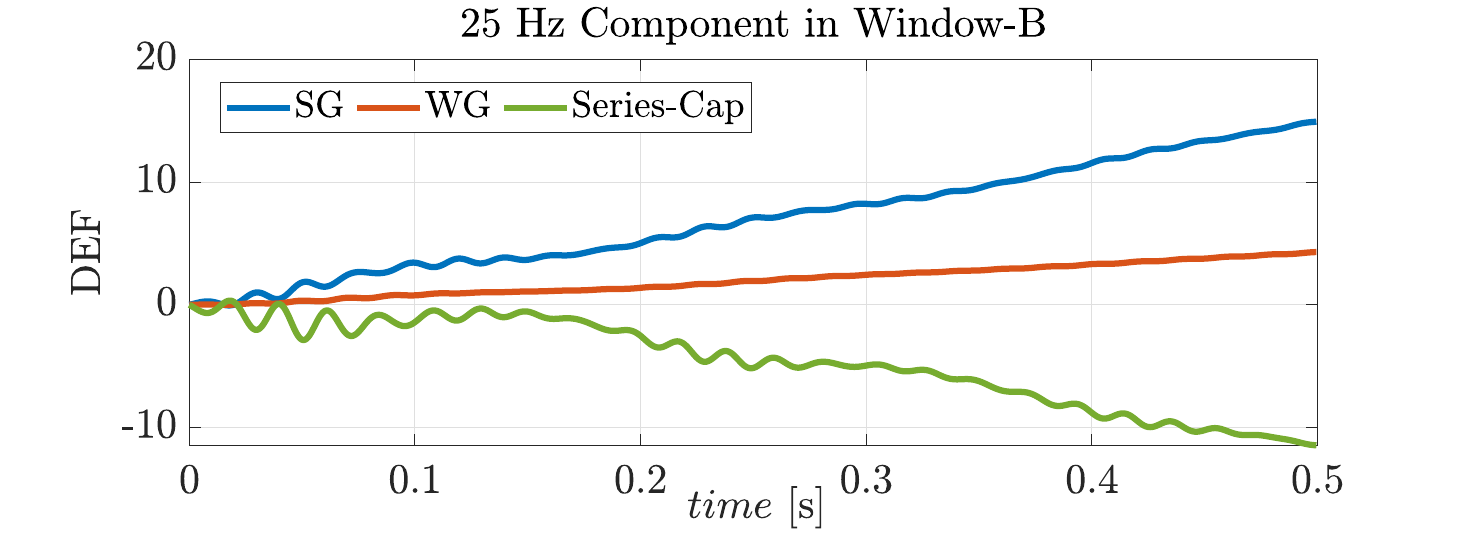}
\includegraphics[width=\linewidth]{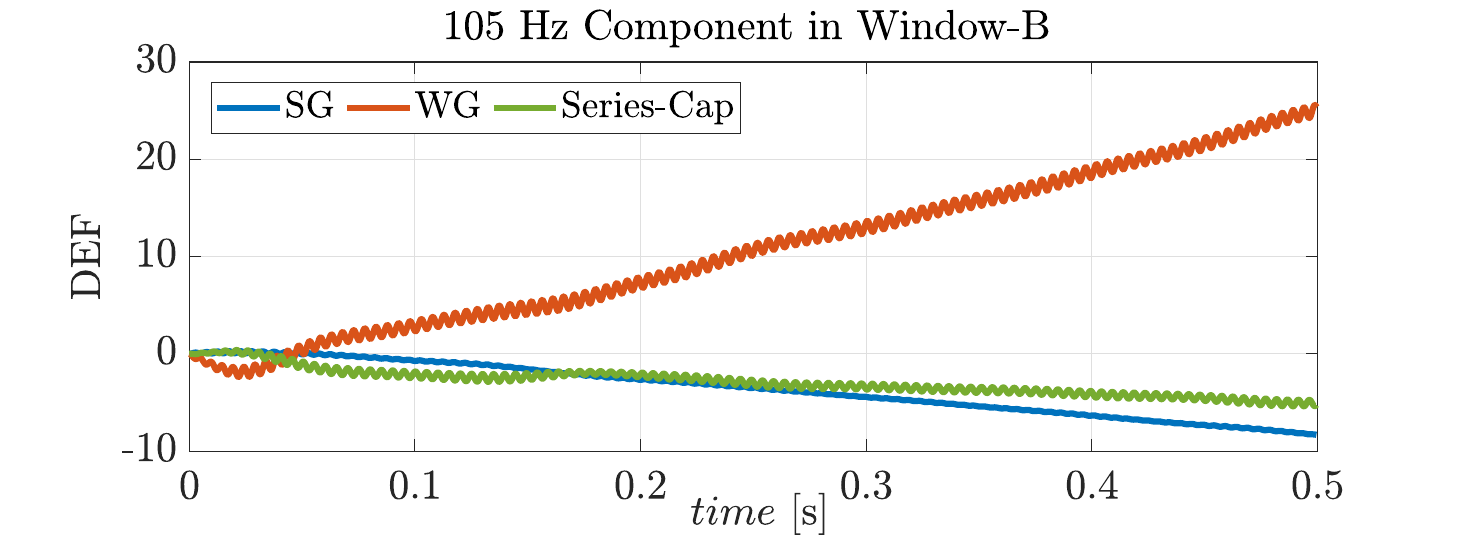}
\vspace{-0.5cm}
\caption{DEF from the grid elements into the grid computed from the data in Window-B between $15–15.5$ s for the (a) 25 Hz and (b) 105 Hz components.}\vspace{-0.3cm}
\label{fig:def_25_after}
\end{figure}

\begin{figure}
\centering
\includegraphics[width=\linewidth]{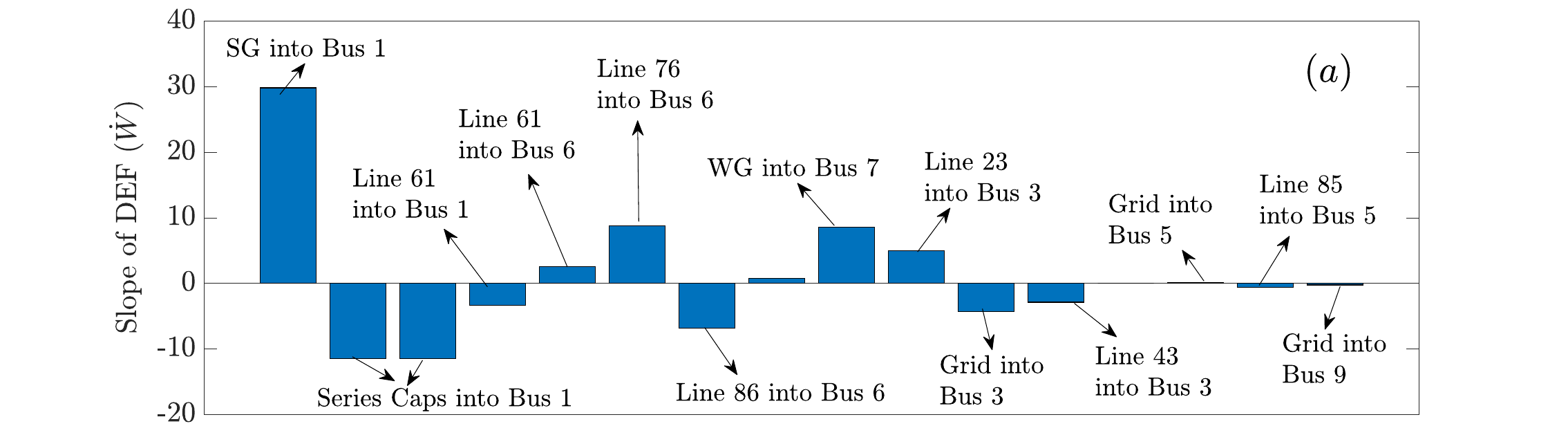}
\includegraphics[width=\linewidth]{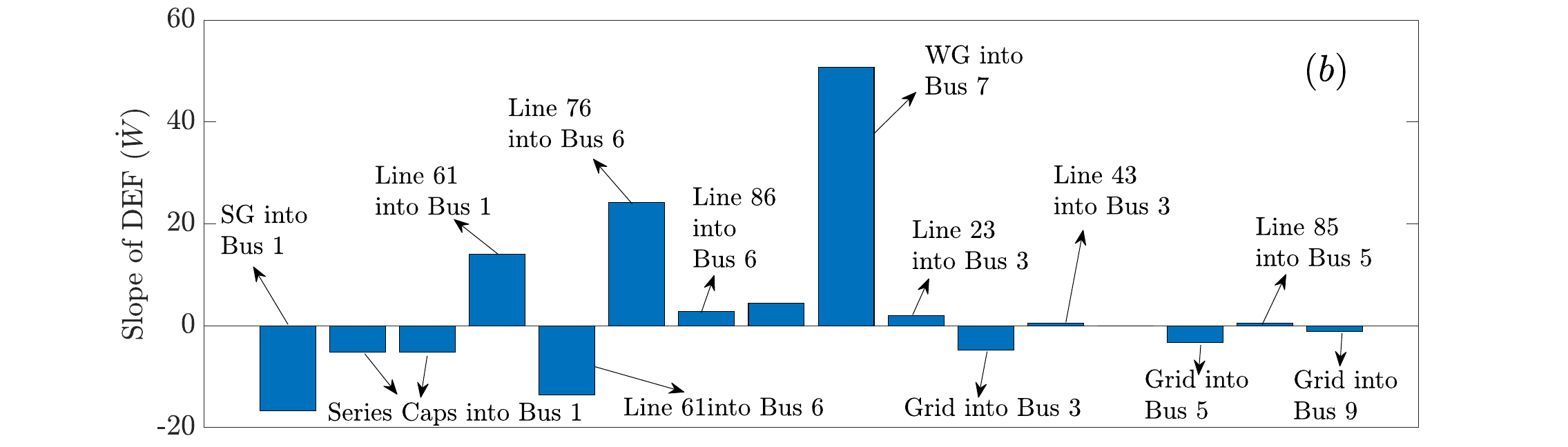}
\vspace{-0.5cm}
\caption{Average slope of DEF from different grid elements calculated for the (a) 25 Hz and (b) 105 Hz components in Window-B.}
\label{fig:slope_def_25_after}\vspace{-0.5cm}
\end{figure}
\begin{figure}
\centering \vspace{-0.4cm}
\includegraphics[width=\linewidth]{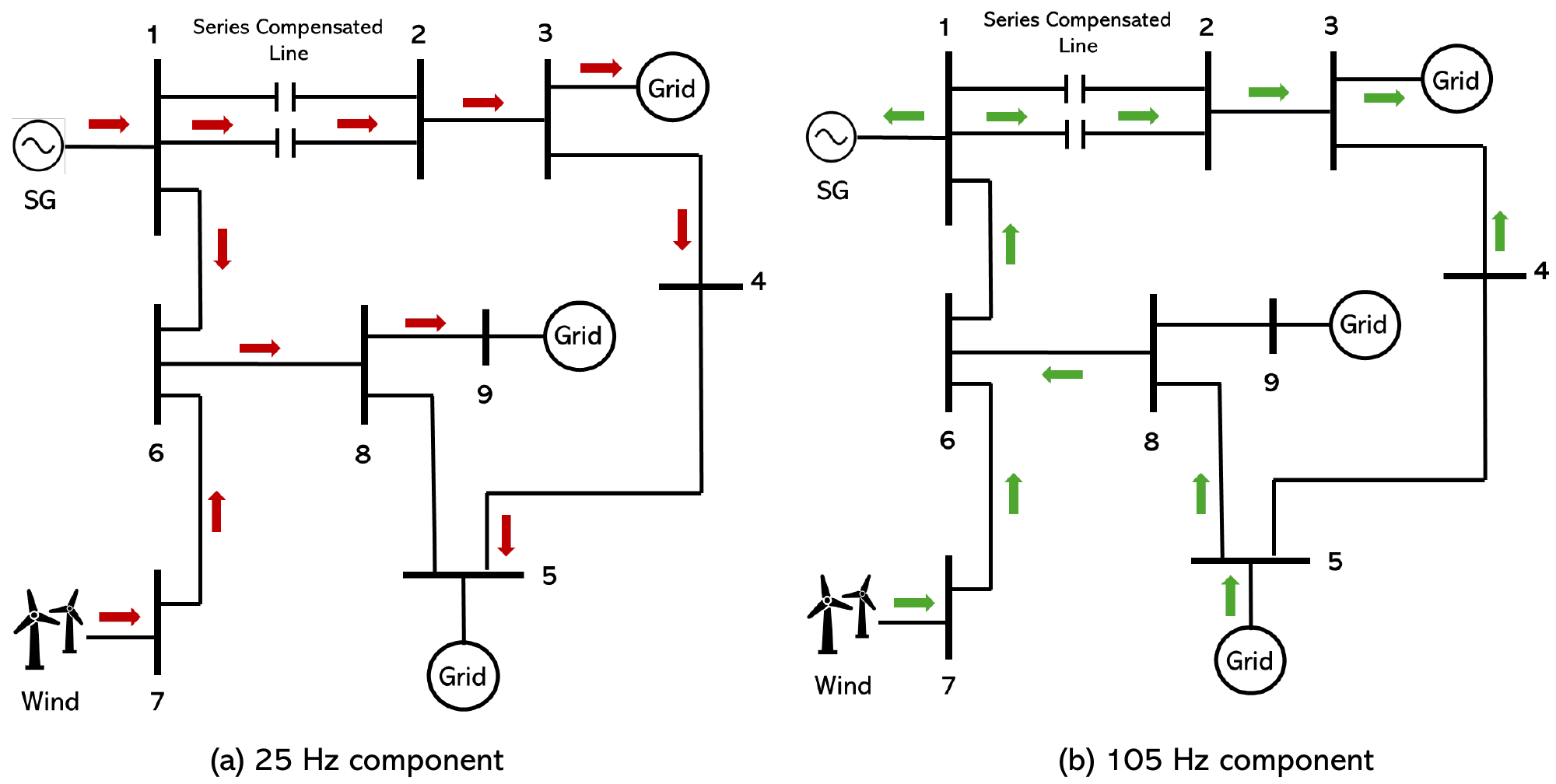}
\caption{Identified SSCI path and oscillation energy flow for the (a) 25Hz and (b) 105Hz components in Window-B.}
\label{fig:path_after}\vspace{-0.5cm}
\end{figure}

In Window-B, DEF is computed separately for the 25 Hz and 105 Hz components (Fig. \ref{fig:def_25_after}). Before detuning, the WG and SG exhibit comparable DEF injections at 25 Hz, indicating similar source behavior. After detuning, the WG’s 25 Hz injection decreases, while a new 105 Hz component emerges. Notably, the SG, which is a source at 25 Hz, becomes a sink at 105 Hz. The 105 Hz oscillation is characterized by the WG as the sole source, with both the SG and the series-compensated line acting as sinks (see Fig. \ref{fig:def_25_after} and Fig. \ref{fig:slope_def_25_after}).

Fig.\ref{fig:path_after} summarizes the identified interaction paths determined from the values in Fig. \ref{fig:slope_def_25_after}. At 25 Hz, oscillation energy propagates between the WG and SG, with dissipation through the series-compensated line. At 105 Hz, the WG acts as the sole source, injecting oscillatory energy into both the SG and the compensated line. Notably, the SG serves as a source for the 25 Hz mode but as a sink for the 105~Hz mode, illustrating how a component’s role in oscillation energy exchange can vary with frequency. This case study demonstrates the capability of DEF to not only locate oscillation sources but also map the corresponding propagation paths across multiple frequencies in a multi-machine network.

\section{Conclusions} 

This paper extends the Dissipative Energy Flow (DEF) framework to identify sources and interaction paths of sub- and super-synchronous control interactions (SSCIs) in inverter-based resource (IBR)-rich grids. The proposed approach operates in the $dq$ domain using dynamic phasors derived from three-phase voltage and current measurements, enabling mode-specific DEF computation for oscillations well above the fundamental frequency. Applied to an EMT test system with synchronous and DFIG-based generation, it successfully localized sources and sinks, mapped propagation paths across multiple frequencies, and captured cases where the same resource acted as a source at one frequency and a sink at another. The approach is easy to implement and interpret, providing actionable insights for targeted control retuning and system reconfiguration. 

\label{sec_conclusions}

\section*{Acknowledgment}
The authors would like to thank Dr. Kaustav Dey at the University of Manitoba, CA, for comments and discussions.
\bibliographystyle{ieeetr}
\bibliography{main}

\end{document}